\documentclass{iaus}
\usepackage{graphicx}

\title[short title of paper] %% give here short title %%
{Grand Challenges in Planetary Nebulae Studies: Binary Evolution and
MHD}

\author[short author list]   %% give here short author list %%
{Adam Frank$^1$}

\affiliation{$^1$Department of Physics and Astronomy, University of
Rochester, \break Rochester, NY 14627, USA \break email:
afrank@pas.rochester.edu}

\pubyear{2006}
\volume{234}  %% insert here IAU Symposium No.
\pagerange{119--126}
\date{?? and in revised form ??}
\jname{Proceedings Title IAU Symposium}
\editors{A.C. Editor, B.D. Editor \& C.E. Editor, eds.}
\begin{document}

\maketitle

\begin{abstract}
We review work on the evolution of planetary nebulae and
proto-planetaries via magneto-rotational mechanisms showing that a
dynamo generated magnetic field can produce the energy and momentum
needed to drive pPN and PNe outflows. Angular momentum
considerations lead to the conclusion that single stars may not be
capable of supporting strong fields for long times. Thus we take the
working hypothesis that most PN may form via binary stars. We
propose that the grand challenge for PN studies is fully
understanding the diverse physical processes at work in binary late
stage evolution including the development of disks, fields and
outflows. 
\end{abstract}

\firstsection % if your document starts with a section,

\section{Introduction: The State of PNe Research}
In recent years the field of Planetary Nebula research has
approached something of a cross-roads.  The impressive success of
radiation transfer, micro-physical modeling of nebular plasma
conditions and 1-D radiation hydro-dynamic models applicable to
spherical nebula have provided confidence that an fairly complete
understanding of parameters within the nebula had been achieved.
This success has also raised the question of the direction for the
field. In the absence of an overarching science question, the field
of PNe studies is in danger of becoming eclipsed by other domains of
interest in astronomy which appear to have more visible and
compelling frontiers.

The impression that PNe is a study of "questions mostly answered" is
wrong however. The last decade has also shown that, in spite of
great progress, our understanding of both meso-physical (Huggins,
these proceedings) and macro-physical (Balick -these proceedings)
nebular characteristics remain far from certain. High resolution
images from platforms like the HST have shown that we still lack a
well tested model which can explain the shapes of PNe and link them,
in a consistent way, to the evolution of the central star.  In this
contribution I review some outstanding issues in pPNe and PNe
shaping and suggest how these provide us with a new way of looking
at PNe studies. The perspective I advocate is that the frontiers of
PNe studies are broad and, most importantly, provide links to some
of the most pressing questions in a variety of other astronomical
environments.

\subsection{Arguments for MHD Launching {\it and} Collimation} The
need for magnetic fields to aid in shaping the nebula has been
discussed before (Balick \& Frank 2002).  The work of Garcia-Segura,
Lopez and others (\cite{ga99}) has demonstrated that including
magnetic fields on nebular scales $R>10^{16} cm$ leads to a wide
variety of morphologies including well collimated jets and point
symmetry. What these models do not address however is the launching
of the wind.

The work of Bujarrabal, Alcolea, Sanchez-Contreras and others makes
"weak field" models untenable and provides one of the most important
challenges to PNe studies.  It also offers a cogent direction
forward for the future. In \cite{buj01} the momentum and energy
budgets for both AGB and pPN outflows where derived. Their
conclusion was that while AGB outflows could be satisfactorily
explained via radiation driving, pPNe outflows showed momentum
"excesses" that where $10^3$ or higher. Such large values of the
ratio of outflow momentum to what could be supplied by photons, $P =
L/c$, could not be accounted for via multiple scattering. Thus
another source of energy is needed to produce the wind. \cite{buj01}
also claimed that the acceleration timescales for the outflows
should be short ($t < 1000$ y).  Thus pPN, and subsequent, PNe may
initially form via an outburst or explosion rather a continuous
wind.

It also noteworthy that the statistics of pPN and young PN
morphologies appear to differ dramatically from fully developed PNe.
The samples of pPNe of Bujarrabal and collaborators and of young PNe
(Sahai \& Trauger 1998) are both dominated by strongly bipolar
outflows. It remains to be seen if this is a observational bias or
not.  If such a distinction between older and younger population
maintains after further study, it will indicate that the early
shaping of PNe occurs via forces which differ as the nebula matures.

\section{Magneto-Rotational Launch (MRL) Models} In magneto-rotational
outflow theory the rotational energy of a central gravitating source
is tapped via an embedded magnetic field to launch material past the
escape speed and create a wind.  At some distance the magnetic
stresses collimate the wind into a narrow jet or broader bipolar
outflow. In works by Blackman, Frank and collaborators MRL models
have been explored in detail. In what follows we briefly review
these works.

{\bf Disk+Star Model:}In Blackman Frank \& Welch 2001 a model was
proposed which assumed a disk forming via disruption of a binary. In
this model both the disk and the rotating core of the central star
produce a wind via magneto-rotational processes. The key expression
related the wind mechanical luminosity to the magnetic luminosity of
the rotator (star or disk)
\begin{equation}
L_w \approx  {\dot M}_w v_{w}^2 \sim L_m = \int ({\bf E} \times {\bf
B}) \cdot d{\bf S} \sim \int_{R_{i}}^{R_{o}} (\Omega R) B_pB_\phi
RdR,
\end{equation}These models were
successful in outlining the mechanisms and rough time-dependencies
of the wind launching and established a basis for a MRL paradigm for
pPNe.

{\bf Disk Only Models:}  In Frank \& Blackman 2004 a version of MRL
launching where centrifugal forces along the poliodial component of
the field dominate initially (a "Fling") was applied to pPNe and PNe
disks. Here scaling relations were derived from equations for such
magneto-centrifugal launching.  From these relations it was shown
that the energy and momentum budgets for pPN could be recovered. In
addition conditions typical of a mature PNe could also be supported
via magneto-centrifugal wind launching.

{\bf Disk Only Models:} In Blackman et al 2001 a single star model
was proposed for the development of collimated pPN outflows.  This
work used a model for an initially rotating $3 M_\odot$ star that
was evolved to the tip of the AGB (Kwaler - private communication)
with the assumption of angular conservation on shells.  An interface
dynamo model was then applied to the resulting rotation profile
$\Omega(R)_*$.  The results showed that a dynamo of sufficient
strength was produced to power an impulsive outflow with energy
matching what was found in pPNe. In this model it was assumed that
the outflow would be initiated as overlaying material was peeled
away by the AGB wind.
\begin{figure}
%\centering \resizebox{7.5cm}{!}{\includegraphics{frank1.eps} }
\caption[]{3-D rendering of field lines and density in magnetic
explosion models.  Left: central regions.  Right: overall
morphology.  Note dominance of toroidal fields.}
\end{figure}

More recently Matt, Frank \& Blackman 2006 have explored the process
of wind launching from an exposed, rotating stellar core in more
detail.  Their simulations tracked a core with an initially dipole
or split monopole field.  The core was surrounded by an envelope in
hydrostatic equilibrium which was not rotating. The differential
rotation between the core and the envelope converted poloidal field
$B_p$ into toroidal field $B_\phi$.  As the field winds up magnetic
pressure gradients $\sim \nabla B_\phi^2$ eventually drive the
envelope beyond the local escape speed (a "Spring" model).  We note
that the key parameter which determined the launch of the outflow
was $\chi = (v_a v_{rot})^{1/2}/v_{esc}$. We found that $\chi > 0.2$
should be the limit for initiating the explosion.

The outflow which forms from such a configuration was strongly
collimated, though an equatorial outflow could also form depending
on the field configuration (Fig 1).  It is important to note that
the outflow is transient producing a kinetic energy dominated "cap"
riding on a Poynting flux dominated jet.  Such Poynting flux
 outflows are the subject of considerable interest in other
domains such as GRBs.  Thus the outflow forms more of a magnetic
explosion than a continuous flow.  Such conditions may be well
suited to describing pPNe.  In a recent set of simulations
(Cunningham \& Frank 2006, in preparation) we have attempted to
model the pPNe, CRL 618 as a collection of bullets impulsively
launched from the source. Our initial results show such a model is
effective at reproducing the gross morphology and kinematics of the
nebula (Fig 2).

\section{Binaries, Dynamos Models and the Future} The importance
of binaries in PNe shaping continues to be a subject of debate.  For
many years Soker (and Livio) has argued that angular momentum
exchange via some form of companion would play a strong role in PNe
evolution (i.e. Soker \& Rappaport 2000: also see DeMarco these
proceedings). Clearly all models involving accretion disks require
the presence of a binary companion. What was not initially
appreciated however was the need for a binary to maintain conditions
for a strong dynamo in an AGB star.  Soker initially made this point
and our own work supports this conclusion. The problem can be
couched in terms of energy considerations as each dynamo cycle
involves the removal of magnetic energy. Thus rotational energy is
continually lost to magnetic field which must then be regenerated.
Recently Nordhaus \& Blackman 2006 have used common envelope models
to map out different evolutionary routes in terms of angular
momentum exchange. In particular, they demonstrate how differential
rotation profiles can be re-energized (from outside-in) via the
influence of a companion. It appears then that so-called single star
MHD models of outflow launching and collimation are untenable. The
influence of a second, orbiting body is always required.

Observations appear to show that the majority of PNe begin life
highly collimated with momentum too high to be driven by radiation
pressure.  Thus we turn to MRL models of PN outflow launching and
collimation only to find that in all cases a companion is needed to
provide conditions to create the field.  {\it \bf In this way it
appears that most PNe form from binaries}.  Our own perspective is
that such a view should be a straw-man hypothesis to focus the
efforts of the PNe community.  Perhaps PNe should be seen as one of
the variety of consequences of binary evolution. Obviously there are
many questions which the hypothesis raises such as the already
observed link between scale-heights and different morphological
populations.  Note also that one way out of the dynamo $=$ binary
conclusion would be for convection to resupply differential rotation
(Blackman 2004). Given the enormous importance of dynamos, MHD
driven outflows and binary evolution across many fields of astronomy
the presence of the issue in our community points the way to a grand
challenge which we could attempt to solve.

\begin{figure}
%\centering \resizebox{7.5cm}{!}{\includegraphics{frank2.eps} }
\caption[]{Simulation of 3 impulsively ejected hypersonic bullets as
a model of CRL 618. Image highlights location of shocks as bullets
propagate down the grid. }
\end{figure}

\begin{acknowledgments}
This work was supported by Spizter Space Telescope theory grant
051080-001, NSF grant AST-0406799 and DOE grant DE-F03-02NA00057

\end{acknowledgments}

\end{document}